\documentclass[12pt]{article}
\usepackage[a4paper]{geometry}
\usepackage{amsmath}
\usepackage{graphicx}
\usepackage{setspace}
\usepackage{hyperref}
\newcommand{\beqn}{\begin{equation}}
\newcommand{\eeqn}{\end{equation}}
\newcommand{\beqna}{\begin{eqnarray}}
\newcommand{\eeqna}{\end{eqnarray}}
\title{Multifractal characterization of gold market: a multifractal detrended fluctuation analysis}
\author{Provash Mali\footnote{E-mail: provashmali@gmail.com} and Amitabha Mukhopadhyay\\
 Department of Physics, University of North Bengal,\\
West Bengal, Darjeeling 734013, India}
\date{}
\begin{document}
\maketitle
\begin{abstract} \noindent
The multifractal detrended fluctuation analysis technique is employed to analyze the time
series of gold consumer price index (CPI) and the market trend of three world's highest
gold consuming countries, namely China, India and Turkey for the period: 1993–July 2013.
Various multifractal variables, such as the generalized Hurst exponent, the multifractal
exponent and the singularity spectrum, are calculated and the results are fitted to the
generalized binomial multifractal (GBM) series that consists of only two parameters.
Special emphasis is given to identify the possible source(s) of multifractality in these series.
Our analysis shows that the CPI series and all three market series are of multifractal nature.
The origin of multifractality for the CPI time series and Indian market series is found due
to a long-range time correlation, whereas it is mostly due to the fat-tailed probability
distributions of the values for the Chinese and Turkey markets. The GBM model series more
or less describes all the time series analyzed here.
\end{abstract}


\section{Introduction}
In general, a fractal is a rough or fragmented geometrical shape that can be subdivided into parts, each of which is (at least approximately) a reduced-size copy of the whole. A fractal system is usually described by a scale invariant parameter called fractal dimension \cite{Mand82}. Many fractals arising in nature have a far more complex scaling relation than simple fractals and require a set of parameters to specify such objects that are known as multifractals.
Several approaches have so far been developed and applied to explore the of fractal properties. For instance, the rescaled adjusted range analysis method was introduced by Hurst \cite{Hurst51,Hurst65} (see also \cite{Feder88}), which he himself applied the method to his hydrological study. Due to the difficulty of the rescaled analysis in capturing long-range correlations of nonstationary series, Peng et al. \cite{Peng94} proposed an alternative approach to analyze the DNA sequences which is known as detrended fluctuation analysis (DFA). Although the DFA method is widely used to determine monofractal scaling properties, it cannot properly describe multi-scale and fractal subsets of time series data. One of the simplest type of multifractal analysis has been developed based on the standard partition function multifractal formalism \cite{Feder88,Bacry01}. This is a highly successful method for the multifractal characterization of normalized and stationary measures, but it does not give the correct result for nonstationary time series. Based on a generalization of the DFA method, Kantelhardt et al. \cite{Kant02} introduced the multifractal detrended fluctuation analysis (MF-DFA) for the multifractal characterization of nonstationary time series. As a remarkable powerful technique, MF-DFA has so far been applied to various fields of stochastic analysis, for instance, in markets return analysis \cite{Matia03,Oswi05,Oswi06,Kwap05,Wang11,Samadder12,Lu13,Mali14}, in geophysics \cite{Kant03,Eva06,Vero11,Shi13,Benicio13}, in biophysics \cite{Liao11,Esen11,Kumar13,Card12}, and also in various branches of basics and applied physics \cite{Zhang08,Wang13,Varotsos10,Igna10,Baroni10,Mur13}.

The study of financial time series has been the focus of intense research by the physics community in the last several years. Nowadays, there are some excellent compilations available on this subject, e.g. \cite{Mantegng1999,Bouchaud2000,Kwapien12}, just to cite some of them. The prime objective of this kind of analysis is to characterize the statistical properties of the time series with the hope that a better understanding of the underlying dynamics could provide useful information to create new models. Moreover, such knowledge might be crucial to tackle relevant problems in finance, such as risk management or the design of optimal portfolios. Henceforth our discussion will be restricted to the time series analysis of gold market in China, Indian, Turkey and the global consumer price index (CPI)\footnote{A consumer price index (CPI) is an estimate as to the price level of consumer goods and services in an economy which is used as a way to estimate changes in prices and inflation. A CPI takes a certain basket of common goods and services, for instance a gallon of gasoline and diesel fuel or an ounce of gold, and tracks the changes in the prices of that basket of goods over time. The gold CPI, according to the World Gold Council \cite{WGC}, is composed of the 5 largest gold consuming country currencies, ranked by and weighted by 3 year average gold demand.}.

Gold being one of the most precious metals is always considered as the safest investment. Presently the fluctuations of gold market seem quite confusing even to the regular traders, and it becomes almost impossible to predict its accurate rise or fall. As we know, over the last 2/3 years gold price increases so rapidly that the gold price nowadays is approximately double of the average rate of 2010. The market gains its highest value of about 1900 USD/ounce in the year 2011, and the recent value is about 1300 USD/ounce. Moreover, the day-to-day variation of the market is also quite remarkable during the last few years. It is now believed that gold is not a commodity anymore, rather a currency which always maintain an inverse relation with the US economy. So it is quite possible that a part of the change in gold price is really just a reflection of a change in the value of the US dollar. Sometimes such change is insignificant and often the opposite is true. Whatever may be the reason, the dynamical nature of the gold market is quite complex, and one needs to study the time series of gold price from all possible directions in order to understand the underlying mechanism.

In this article we apply the MF-DFA technique to characterize the time series of the gold CPI and the gold market in China, India and Turkey during the period 1993--July 2013. According to the World Gold Council, these three countries are the world's major gold consuming countries with a combined consumption of about $70\%$ of the total demand. The individual consumptions of these countries are: China $33\%$, India $28\%$ and Turkey $9\%$. Hence, it is expected that the CPI of gold is mainly governed by the markets of these three countries. In order to visualize the recent market pattern, we separately analyze the series of about the last three years period--from 2010 to June 2013. Various parameters related to multifractality of the studied time series are computed and are fitted to the two-parameter generalized binomial multifractal (GBM) model \cite{Feder88,Kant03}. Special emphasis is given to identify the probable origin(s) of multifractality in these series. For this purpose we analyze a randomly shuffled series and a surrogate series corresponding to each of the original series.  
The article is organized as follows: in Section 2 the MF-DFA methodology is described along with the outlines of the GBM model. In Section 3 we describe the data and the results of our analysis, and the article is summarized in Section 4.

\section{MF-DFA methodology}
Though nowadays the MF-DFA technique has become a standard tool of time series analysis, for the sake of completeness we provide in this section a brief description of the method which is followed by the outlines of the GBM model used to compare the empirical data.

Let $\{x_k:~k=1,~2, \dots,N\}$ be a time series of length $N$. The MF-DFA procedure consists of the following five steps:\\
{\em Step 1}: Determine the profile
\beqn
Y(i) = \sum_{k=1}^{i} [x_k - \left<x\right>],~~ i=1,~2,\dots, N,
\label{eq:Profile}
\eeqn
where $\left<x\right> = (1/N)\sum_{k=1}^N x_k$ is the mean value of the analyzed time series.\\
{\em Step 2}: Divide the profile $Y(i)$ into $N_s = int(N/s)$ non-overlapping segments of equal length $s$. One has to choose the $s$ value depending upon the length of the series. In the case, the length $N$ is not a multiple of the considered time scale $s$, the same dividing procedure is repeated starting from the opposite end of the series. Hence, in order not to disregard any part of the series, usually altogether $2N_s$ segments of equal length are obtained.\\
{\em Step 3}: Calculate the local trend for each of the $2N_s$ segments. This is done by a least-square fit of the segments (or subseries). Linear, quadratic, cubic or even higher order polynomial may be used to detrend the series, and accordingly the procedure is said to be the MF-DFA1, MF-DFA2, MF-DFA3, $\dots$ analysis. Let $y_p$ be the best fitted polynomial to an arbitrary segment $p$ of the series. Then determine the variance
\beqn
F^2(p,s) = \frac{1}{s}\sum_{i=1}^{s}\left\{Y[(p -1)s+i] - y_{p}(i) \right\}^2
\eeqn
for $p=1, \dots N_s$, and for $p = N_s+1, \dots, 2N_s$ it is given as,
\beqn
F^2(p,s) = \frac{1}{s}\sum_{i=1}^{s}\left\{Y[N-(p -N_s)s+i] - y_{p}(i) \right\}^2.
\eeqn
{\em Step 4}: Define the $q$th order MF-DFA fluctuation function
\beqn
F_q(s) = \left\{ \frac{1}{2N_s} \sum_{p=1}^{2N_s}[F^2(p,s)]^{q/2} \right\}^{1/q}
\label{eq:Fq1}
\eeqn
for all $q \neq 0$ and for $q=0$ the above definition is modified to the following form
\beqn
F_q(s) = \exp\left\{ \frac{1}{4N_s} \sum_{p=1}^{2N_s}\ln [F^2(p,s)] \right\}.
\label{eq:Fq2}
\eeqn
{\em Step 5}: Then the scaling behavior of the fluctuation functions is examined for several different values of the exponent $q$.

If the series $\{x_k\}$ possess long-range (power-law) correlation, $F_q(s)$ for large values of $s$ would follow a power-law type of scaling relation like
\beqn
F_q(s) \sim s^{h(q)}.
\label{eq:Scaling}
\eeqn
In general, the exponent $h(q)$ depends on $q$ and is known as the generalized Hurst exponent. For a stationary time series $h(2) =H$ -- the well known Hurst exponent \cite{Kant03}. For a monofractal series on the other hand, $h(q)$ is independent of $q$, since the variance $F^2(p,s)$ is identical for all the subseries and hence Eqns. \eqref{eq:Fq1} and \eqref{eq:Fq2} yield identical values for all $q$. Note that the fluctuation function $F_q(s)$ can be defined only for $s \geq m+2$, where $m$ is the order of the detrending polynomial. Moreover, $F_q(s)$ is statistically unstable for very large $s~(\geq N/4)$. If small and large fluctuations scale differently, there will be a significant dependence of $h(q)$ on $q$. For positive values of $q$, $F_q(s)$ will be dominated by the large variance which corresponds to the large deviations from the detrending polynomial, whereas for negative values of $q$, major contributions in $F_q(s)$ arise form small fluctuations from the detrending polynomial. Thus, for positive/negative values of $q$, $h(q)$ describes the scaling behavior of the segment with large/small fluctuations.

One can easily relate the $h(q)$ exponent with the standard multifractal exponent, such as the multifractal (mass) exponent $\tau(q)$. Consider that the series $\{x_k\}$ is a stationary and normalized one. Then the detrending procedure in step 3 of the MF-DFA methodology is not required, and the variance of such series $F^2_N$ is given by
\beqn
F^2_N(p,s) = \{Y(p s) - Y[(p-1)s]\}^2.
\label{eq:Vari2}
\eeqn
Then the fluctuation function and its scaling law are given by
\beqn
F_q(s) = \left\{\frac{1}{2N_s} \sum_{p = 1}^{2N_s}|Y(p s) - Y[(p-1)s]|^q \right\}^{1/q} \sim s^{h(q)}.
\label{eq:Fq2}
\eeqn
Now if we assume that the length of the series $N$ is an integer multiple of the scale $s$, then the above relation can be rewritten as,
\beqn
\sum_{p=1}^{N/s} |Y(p s) - Y[(p -1)s]|^q \sim s^{qh(q) -1}.
\label{eq:Fq3}
\eeqn
In the above relation the term under $|\cdot|$ is nothing but sum a of $\{x_k\}$ within an arbitrary $p$th segment of length $s$. In the standard theory of multifractals it is known as the box probability $\mathcal P(s,p)$ for the series $x_k$. Hence,
\beqn
\mathcal P(p,s) \equiv \sum_{k=(p-1)s+1}^{p s} x_k = Y(p s) - Y((p -1)s).
\label{eq:BoxPro}
\eeqn
The multifractal scaling exponent $\tau(q)$ is defined via the partition function $Z_p(s)$
\beqn
Z_{\mathcal P}(s) \equiv \sum_{p=1}^{N/s} |\mathcal P(p,s)|^q \sim s^{\tau(q)},
\label{eq:Z}
\eeqn
where $q$ is a real parameter.
From Eqns.~\eqref{eq:Fq3}-\eqref{eq:Z} it is clear that the multifractal exponent $\tau(q)$ is related to $h(q)$ through the following relation:
\beqn
\tau(q) = qh(q) - 1.
\label{eq:tau}
\eeqn
Knowing $\tau(q)$ one can calculate the most important parameter of a multifractal analysis -- the multifractal singularity spectrum (also called the spectral function) $f(\alpha)$ which is related to $\tau(q)$ through a Legendre transformation \cite{Feder88,Peitgen92}: $\alpha = \partial \tau(q)/\partial q$, and is given by
\beqn
f(\alpha) = q \alpha - \tau(q).
\label{eq:fa}
\eeqn
Here $\alpha$ is the singularity strength or Holder exponent. Importance of the singularity spectrum in the theory of multifractals is that, the width of the spectrum is a direct measure of the degrees of multifractality. For a monofractal object it turns out to be a delta function at the corresponding $\alpha$.  \\

The binomial multifractal series of $N=2^{n_{\max}}$ numbers is defined as,
\beqn
x_k = a^{n(k-1)}(1-a)^{n_{\max} - n(k-1)},
\label{eq:bin}
\eeqn
with $k=1, \dots, N$, $n(k)$ is the number of digits equal to 1 in the binary representation of the index $k$ and $0.5 < a < 1$ is a parameter. A generalization of this series with two positive valued parameters (say, $a$ and $b\equiv (1-a)$) is given as,
\beqn
x_k = a^{n(k-1)}b^{n_{\max} - n(k-1)}.
\label{eq:binomial}
\eeqn
The above generalized form of the binomial multifractal series \eqref{eq:bin} has been used in \cite{Kant03} to characterize the river runoff data. In the text it is said to be the generalized binomial multifractal (GBM) series. With this generalization one can easily derive the expressions for the multifractal observables $h(q)$ and $\tau(q)$ of the following form:
\beqna
h(q)    &=& \frac{1}{q} - \frac{\ln [a^q+b^q]}{q\ln 2}. \label{eq:hq} \\
\mbox{and}~~\tau(q) &=& - \frac{\ln [a^q+b^q]}{\ln 2}, \label{eq:tauq} 
\eeqna
Note that for $a=b$, $h(q) = -\ln a/\ln2$ is independent of $q$ and $\tau(q) \sim q \ln a/\ln 2$ is a linear function of $q$, i.e. for $a=b$ the series \eqref{eq:binomial} reduces to a monofractal series. Then obviously the quantity $|a-b|$ is a measure of the strength of multifractality. Quantitatively, the strength parameter $\Delta \alpha$ is defined by the difference of the asymptotic values of $h(q)$, i.e., $\Delta \alpha \equiv h(-\infty) - h(+\infty) = (\ln a -\ln b)/\ln2$, which is nothing but the width of the singularity spectrum at $f(\alpha)=0$.

\section{Results and discussion}
\begin{figure}[t]
\centering
\includegraphics[width=6.5in,height=5.5in]{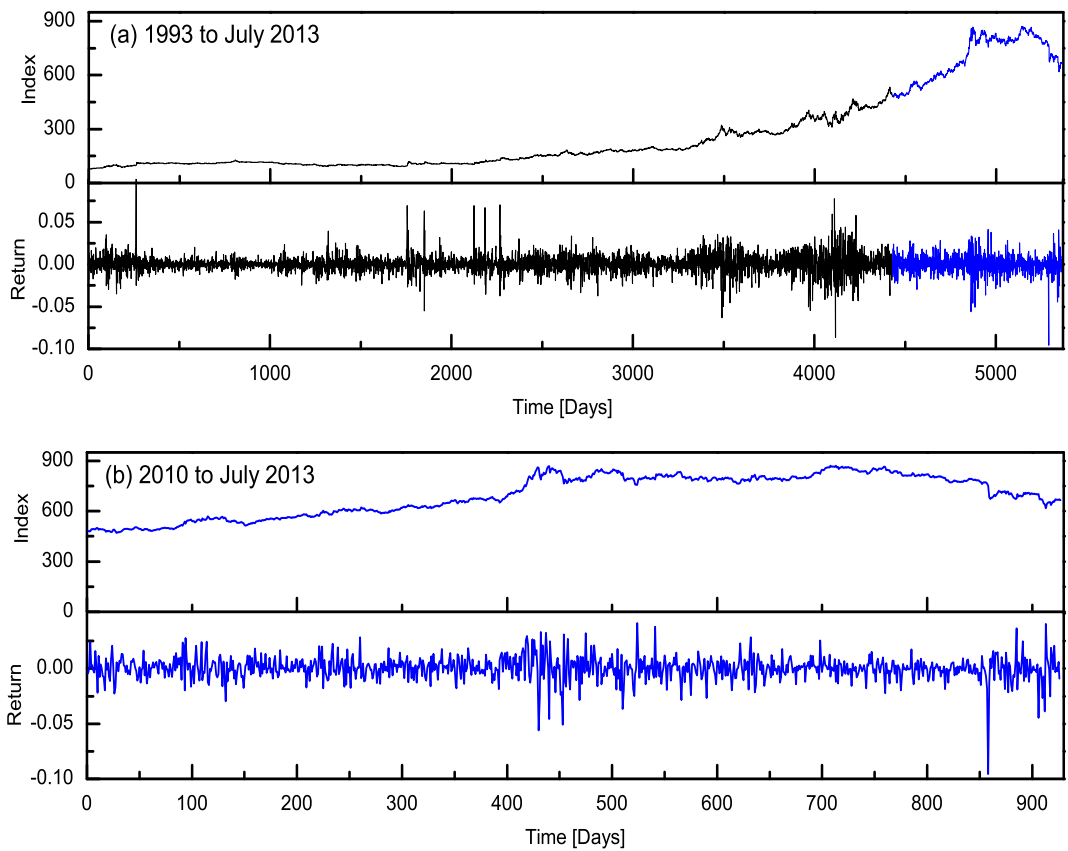}
\caption{(a) Time series of the gold consumer price index and the corresponding return for the time period 1993--July 2013, (b) same as (a) but for the time period 2010--July 2013.}
\label{Fig1}
\end{figure}
\begin{figure}
\centering
\includegraphics[width=5.5in,height=4in]{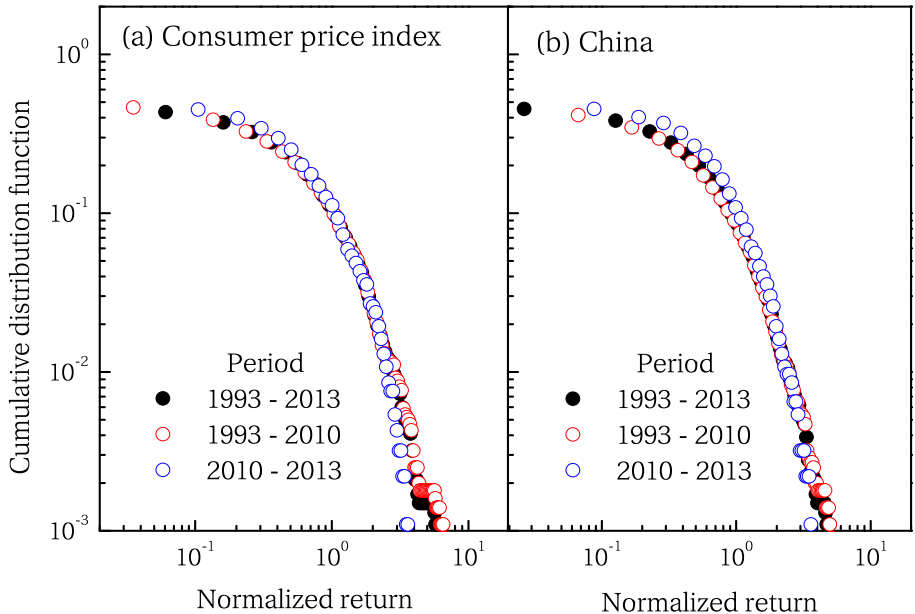}
\caption{Cumulative distribution function of the normalized returns of gold price indices for three different time periods: (a) global consumer price index and (b) time series of China.}
\label{Fig1B}
\end{figure}
 The gold market data used in this article are taken from the database of the World Gold Council \cite{WGC}. The logarithmic difference between two successive trading days, also known as return: $R(T) = \ln P(T+1) - \ln P(T)$, is used as the original series for this analysis. To illustrate the nature of the series we show in Fig.~\ref{Fig1}(a) the CPI time series (upper panel) along with its returns (lower panel). The last section of these series shown in blue color are for the period of 2010--July 2013 and is magnified in Fig.~\ref{Fig1}(b). In the text the first section, i.e. from 1993 to 2010, of the series is identified as series of period-I and the last section (from 2010 to July 2013) is identified as period-II. Note that, the time periods are explicitly mentioned in the diagrams. Figure \ref{Fig1} demonstrates how the global gold market evolves with time. 
 In Fig.~\ref{Fig1B} we show the log-log plots of the cumulative distribution function (CDF) of normalized returns of gold price indices for (a) the CPI and (b) the market series in China. The CDFs for the other series analyzed here are more or less similar to \ref{Fig1B}(b) and hence are not shown here. However, the tail exponent, ($\zeta$), gives the critical order of divergence of moments, is extracted from power-law regression for all the series. The $\zeta$ values are all found to vary from 3.04 to 3.5 for the long series, whereas for the series of period-II (2010-2013) the values are as high as 5.0 (for the CPI series).

\begin{figure}[t]
\centering
\includegraphics[width=6.5in,height=5in]{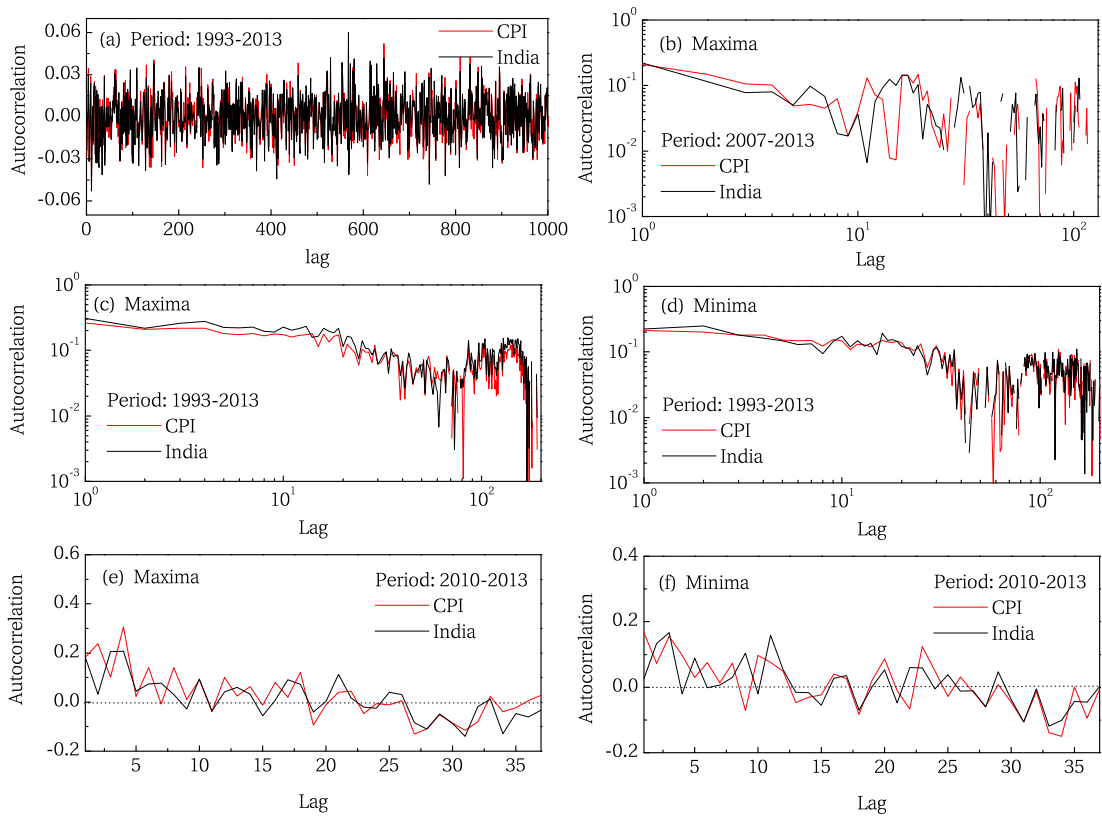}
\caption{Autocorrelation functions for the gold consumer price index (CPI) series compared with the price time series in Indian market. (a) Includes total series for the period 1993 - July 2013, which is totally uncorrelated. (b)--(f) are drawn for the maxima/minima returns in windows of length 5 for several different time periods as mentioned in the diagrams. Notice the signature of long-range correlation in (b)--(d). The correlation function in (b) for a period of about 6 years captures a very weak correlation, whereas (e) and (f) do not possesses any noticeable correlation pattern.}
\label{Corr}
\end{figure}
 The first indication for a power-law type of CDF of market return fluctuations can be traced back to \cite{Lux96} and explicitly in stock markets \cite{Gopi98, Gopi99} the power-law is found to be inverse cubic. Gabaix et al. \cite{Gabaix06} have given a theoretical  interpretation of the above empirical observations. In the case of gold market returns for the period 1968 -- 2010 the tail exponent is estimated to be $\approx3$, i.e. the CDF follows inverse cubic power-law \cite{Bolgorian11}. The effect of faster departure from the inverse cubic power-law in the more recent financial data, similar to the one shown in the present manuscript for the  gold price time series in the period 2010--2013, is also shown in \cite{Drozdz03,Drozdz07,Drozdz10}. Recently, Rak et al. \cite{Rak13} indicate a possible more general applicability of the concept of Gabaix et al. \cite{Gabaix06} to the situations when the price fluctuations depart from the inverse cubic power-law.

Long term correlated records $\{x_i:i=1,2, \dots N \}$ with zero mean and unit variance are characterized by an autocorrelation function $C_x(s) = \left< x_i x_{i+s} \right> \equiv 1/(N-s) \sum_{i=1}^{N-s}x_i x_{i+s}$. If $\{x_i\}$ are uncorrelated, $C(s)$ is zero for $s>0$. Short-range correlations of $\{x_i\}$ are described by $C(s)$ declining exponentially: $C(s) \sim \exp(−s/s_{\times} )$ with a decay time $s_{\times}$. For long-range correlations $C(s)$ declines as a power-law: $C(s) \sim s^{\gamma}$, where the correlation exponent $\gamma$ is between 0 and 1. A direct calculation of $C(s)$ is usually not preferable because of the underlying trends of the series $\{x_i\}$ and its noise factor. 
To test for long-term correlations, it is useful to employ the detrended fluctuation analysis
 (DFA) in order to extract the Hurst ($H$) exponent, which is related to the $\gamma$ exponent as: $H = 1 - \gamma/2$ \cite{Peng94}. However, as a preliminary estimate of the nature of correlation present in the gold price time series data analyzed here, we evaluate the autocorrelation functions for the series. In Fig. \ref{Corr} we illustrate the autocorrelation functions for the consumer price index time series as well as for the Indian market time series. We observe that the daily returns of gold indices, irrespective of the market, are uncorrelated (Fig. \ref{Corr}(a)). But the maxima and minima of the individual market returns over a reasonable time duration (at least 5-6 years) in windows of length $R=5$ days (1 trading week) separately exhibit long-range correlations (Fig. \ref{Corr}(b)-(f)). Therefore, we do not expect long-range correlation in the series of period 2010-2013. Note that the series of maximum (minimum) returns corresponding to each of the original ones are constructed by selecting maximal (minimal) values of daily returns over each interval of length $R$. Obviously, both sub-sequences contain $R$ times fewer points than the original ones. Another importance of the market return autocorrelation function is that, it can serve as a proxy for market liquidity, though the later is an elusive concept, and hence is not a direct observable. Moreover, market liquidity has several dimensions of measurement (e.g., see \cite{Amihud05}). So a rigorous estimate of the gold market liquidity is beyond the scope of the present manuscript, which will otherwise be confined to the short/long-range correlations of gold market returns and its (multi)fractal nature.

\begin{figure}[t]
\centering
\includegraphics[width=6.8in,height=7in]{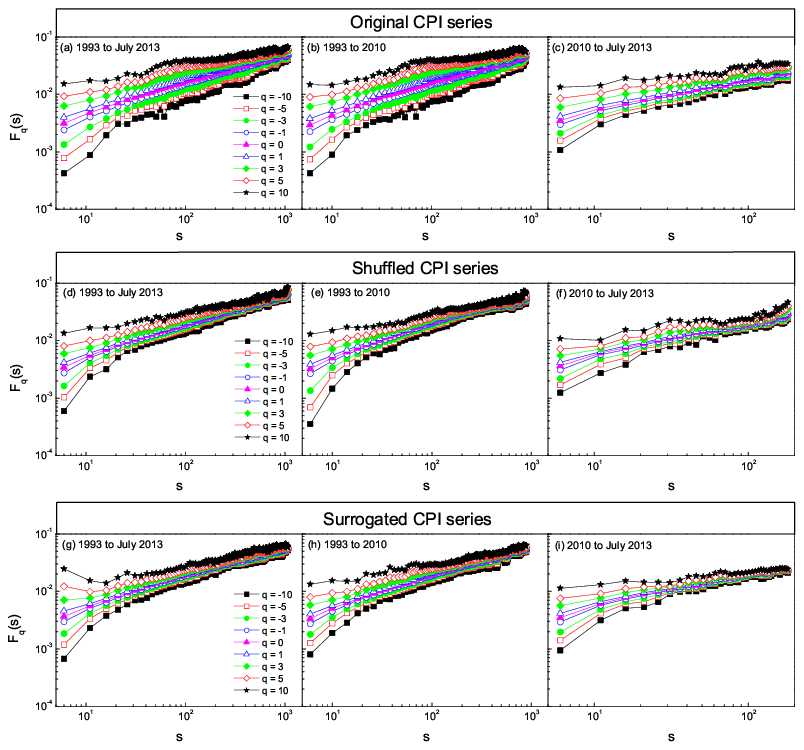}
\caption{The MF-DFA2 fluctuation functions $F_q(s)$ for the consumer price index series of gold plotted with scale $s$ for three different time duration: (a) 1993--July 2013, (b) 1993--2010 and (c) 2010--July 2013. Predictions from the shuffled series (middle panel) and the surrogated series (bottom panel) corresponding to the original ones are also shown.}
\label{Fig2}
\end{figure}
\begin{figure}[t]
\centering
\includegraphics[width=6.75in,height=7in]{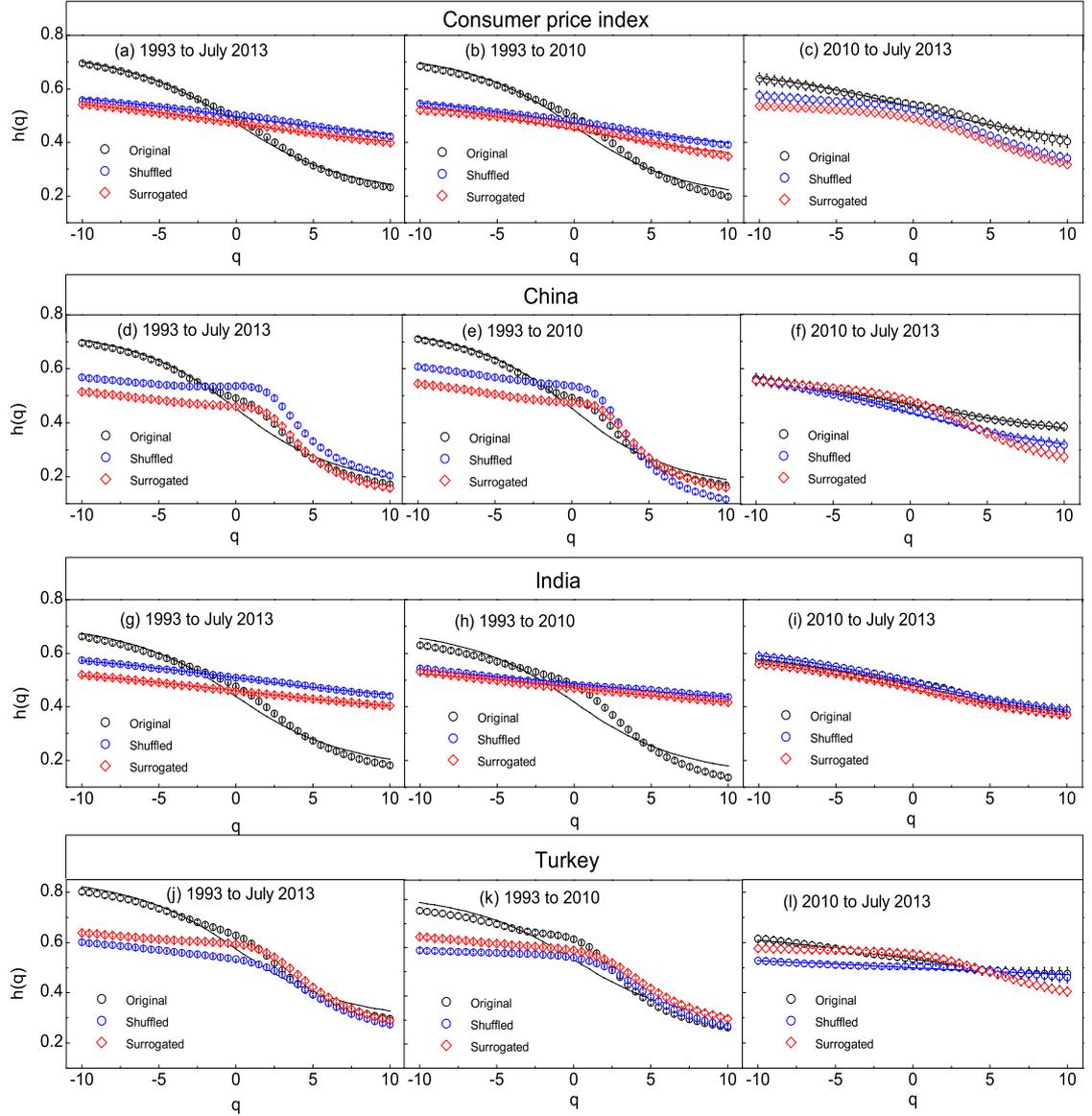}
\caption{The generalized Hurst exponent $h(q)$ spectra for the analyzed time series. The lines correspond to the GBM series \eqref{eq:binomial}, yielding Eqn.~\eqref{eq:hq}. The best fit parameter values are quoted in Table \ref{Table1}.}
\label{Fig3}
\end{figure}
\begin{figure}[t]
\centering
\includegraphics[width=6.75in,height=7in]{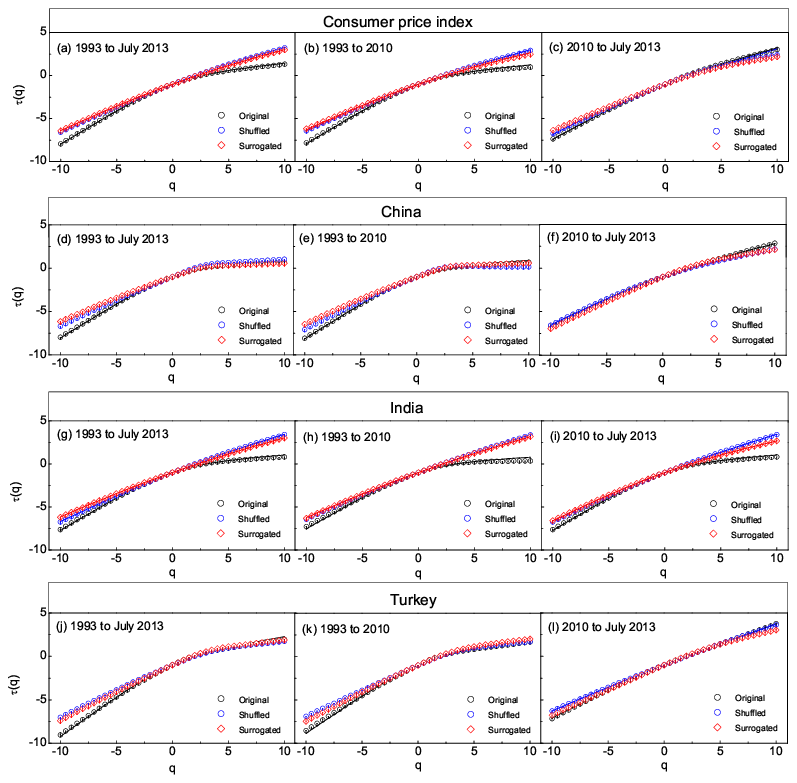}
\caption{The multifractal $\tau(q)$ exponent spectra for the analysed time series. The lines correspond to the GBM series \eqref{eq:binomial}, yielding Eqn.~\eqref{eq:tauq} with the same set of parameters as optimized for the $h(q)$ versus $q$ data points (see Table \ref{Table1}).}
\label{Fig4}
\end{figure}
We calculate the MF-DFA2 fluctuation functions $F_q(s)$ for all the series under consideration and over a wide range of the exponent $q:$ -10 to $+10$ in steps of 0.5. The scaling behaviors of some of the fluctuation functions calculated from the CPI series are illustrated in Fig.~\ref{Fig2}. Separate diagrams are drawn for the total series, period-I and period-II. Here the time scale $s$ is varied from 6 to $N/5$, where $N$ is the series length. In the same figure we also show the $F_q(s)$ functions generated from the randomly shuffled series (middle panel) as well as from the surrogated series (lower panel) corresponding to the original ones. The fluctuation functions calculated for the other series under consideration also follow almost identical scale dependence as shown in the figure, and hence those diagrams are not included here. It is noticed that the multifractal results reasonably vary between the first and second order detrending, but the change is almost insignificant between the second and third or higher order detrending. Hence, the quadratic (second order) detrending is considered for this analysis. It is clear from Fig.~\ref{Fig1} that the scaling behavior of $F_q(s)$ for all three time intervals as well as their shuffled and surrogated counterparts nicely follow the scaling-law \eqref{eq:Scaling} for $s \geq 10$.

It is a known fact that there are two different types of multifractality may exist in a time series data, namely (i) multifractality due to long-range time correlations of the small and large fluctuations and (ii) multifractality due to a fat-tailed probability distribution function of the values in the series. The first kind of multifractality can be removed by random shuffling of the given series and the corresponding shuffled series will exhibit monofractal scaling. Obviously, the probability distribution will not alter by random shuffling and hence the multifractality of the second kind will remain intact. If a given series contains both kinds of multifractality, the corresponding shuffled series will exhibit weaker multifractality than the original one. On the other hand, the surrogate (phase randomization) analysis is an empirical technique of testing nonlinearity for a time series. The aim is to test whether the dynamics are consistent with linearly filtered noise or a nonlinear dynamical system \cite{Uhlenbeck1,Rath12}. The basic idea of the surrogate data method is to first specify some kind of linear stochastic process that mimics ``linear properties'' of the original data. If the predictions (statistics) of the original data are significantly different from those of surrogate series, we may consider the presence of some higher order temporal correlations, i.e., the presence of dynamic nonlinearities. In this analysis we use the the null hypothesis, amplitude-adjusted Fourier transform (AAFT) algorithm \cite{Theiler92} to generate the surrogate series.

\begin{table}
\caption{The best fitted parameter ($a \mbox{ and } b$) values of the generalized binomial multifractal series \eqref{eq:binomial} and the multifractal strength parameter: $\Delta \alpha = (\ln a - \ln b)/\ln2$. The missing data indicate that the binomial series is unable to describe the corresponding time series.}
\centering
\begin{tabular}{lllllll}
\hline \hline
Origin &&Series period &Series type & $a$ & $b$ &$\Delta \alpha$\\
\hline
&&            &Original  &   0.575$\pm$0.008 &  0.904$\pm$0.013 & 0.652$\pm$0.020\\
&&1993 to July 2013  &Shuffled  & 0.639$\pm$0.012 &  0.786$\pm$0.015 & 0.297$\pm$0.027\\
&&            & Surrogated&  0.648$\pm$0.011 &  0.802$\pm$0.014 & 0.306$\pm$0.025\\
\cline{3-7}
&&            &Original  &   0.576$\pm$0.008 &  0.917$\pm$0.013 & 0.670$\pm$0.021\\
CPI &&1993 to 2010  &Shuffled  &  0.645$\pm$0.011 &  0.805$\pm$0.014 & 0.319$\pm$0.025\\
&&            & Surrogated&   0.651$\pm$0.011 &  0.826$\pm$0.013 & 0.344$\pm$0.023\\
\cline{3-7}
&&            &Original  &   0.632$\pm$0.011 &  0.805$\pm$0.014 & 0.351$\pm$0.025\\
&&2010 to July 2013  &Shuffled  & | &  | & |\\
&&            & Surrogated&       | &  | & |\\
\hline 
&&            &Original  &   0.572$\pm$0.008 &  0.933$\pm$0.014 & 0.706$\pm$0.021\\
&&1993 to July 2013  &Shuffled  & | &  | & | \\
&&            & Surrogated&       | &  | & | \\[0.5ex]
\cline{3-7}
&&            &Original  &   0.568$\pm$0.008 &  0.938$\pm$0.013 & 0.724$\pm$0.021\\
China &&1993 to 2010  &Shuffled  &  | &  | & |\\
&&            & Surrogated&         | &  | & |\\[0.5ex]
\cline{3-7}
&&            &Original  &   0.640$\pm$0.011 &  0.816$\pm$0.013 & 0.351$\pm$0.023\\
&&2010 to July 2013  &Shuffled  &   0.639$\pm$0.009 &  0.852$\pm$0.012 & 0.416$\pm$0.021\\
&&            & Surrogated&   | &  | & |\\
\hline 
&&            &Original  &   0.585$\pm$0.008 &  0.929$\pm$0.013 & 0.667$\pm$0.020\\
&&1993 to July 2013  &Shuffled  & 0.635$\pm$0.013 &  0.678$\pm$0.016 & 0.291$\pm$0.028\\
&&            & Surrogated&  0.662$\pm$0.012 &  0.798$\pm$0.014 & 0.272$\pm$0.026\\[0.5ex]
\cline{3-7}
&&            &Original  &   0.593$\pm$0.009 &  0.946$\pm$0.014 & 0.674$\pm$0.021\\
India &&1993 to 2010  &Shuffled  & 0.654$\pm$0.013 & 0.779$\pm$0.016 & 0.252$\pm$0.029\\
&&            & Surrogated&   0.659$\pm$0.013 &  0.790$\pm$0.015 & 0.262$\pm$0.027\\[0.5ex]
\cline{3-7}
&&            &Original  &   0.630$\pm$0.011 &  0.816$\pm$0.013 & 0.374$\pm$0.024\\
&&2010 to July 2013  &Shuffled  &   0.623$\pm$0.010&  0.814$\pm$0.013 & 0.385$\pm$0.024\\
&&            & Surrogated&   0.636$\pm$0.010 &  0.824$\pm$0.013 & 0.372$\pm$0.019\\
\hline 
&&            &Original  &   0.528$\pm$0.010 &  0.853$\pm$0.016 & 0.629$\pm$0.026\\
&&1993 to July 2013  &Shuffled  & | &  | & |\\
&&                   & Surrogated&  | &  |  &|   \\[0.5ex]
\cline{3-7}
&&            &Original  &   0.541$\pm$0.011 &  0.862$\pm$0.017 & 0.671$\pm$0.028\\
Turkey &&1993 to 2010  &Shuffled  & | & | & |\\
&&            & Surrogated&   | &   | & |\\[0.5ex]
\cline{3-7}
&&            &Original  &   0.618$\pm$0.012 &  0.771$\pm$0.015 & 0.319$\pm$0.028\\
&&2010 to July 2013  &Shuffled  &   0.663$\pm$0.017&  0.754$\pm$0.020 & 0.185$\pm$0.038\\
&&            & Surrogated&     | &  | & |\\
\hline\hline
\end{tabular}
\label{Table1}
\end{table}

The generalized Hurst exponent $h(q)$ is extracted from straight line fit to the log-log data of $F_q(s)$ versus $s$. We fit straight line in the $50 \leq s \leq 800$ region for the total and the period-I series, whereas the range chosen for the period-II series is $10 \leq s \leq 60$. Note that, the fluctuation functions are found to obey the scaling law \eqref{eq:Scaling} better in the mentioned scale regions. The order ($q$) dependence of the $h(q)$ exponents is shown in Fig.~\ref{Fig3} for the CPI series as well as for the price time series in China, India and Turkey. The lines in the diagrams represent the best fits of the GBM model (whenever possible) to the data points. The best fit parameter values are given in Table \ref{Table1}. The missing values in the table (and line in the figure) denote that the GBM model cannot describe the corresponding time series. The figure shows that, irrespective of the series time duration, all the $h(q)$ spectra representing the original series are order dependent, though the dependency is weaker for the series of period-II. Moreover, the spectra obtained from the original series are more or less fitted to the GBM model, yielding Eqn.~\eqref{eq:hq}. hOWEVER, in some cases there is noticeable deviation between the empirical values and the model around $q \sim 0$ regions. The shuffled series, on the other hand, produces $h(q)$ spectra that are quite different from their original counterparts. In all cases but the CPI series of period-II, the total and period-I series of China and Turkey, we observe a very weak order dependence of $h(q)$ and the spectra are well described by the GBM model. In view of our previous discussion, the origin of multifractality in those series are both long-range temporal correlation and fat-tailed probability distribution of the variables. In the exceptional cases, the shuffled series produces $h(q)$ spectra that are apparently consistent with an uncorrelated multifractal series with power-law distribution function: $P(x) = \alpha x^{-(\alpha + 1)}$ for $1 \leq x < \infty$ and $\alpha > 0$. For the above series the generalized Hurst exponent is given as \cite{Kant02}: $h(q) \sim 1/q$ for $q > \alpha$ and $\sim 1/\alpha$ for $q \leq \alpha$, whereas $\tau(q) = q/\alpha - 1$ for $q \leq \alpha$ and $\tau(q)=0$ for $q > \alpha$. Note that the multifractality of the above kind is originated purely from a fat-tailed probability distribution function. According to refs. \cite{Nakao00,Drozdz09}, the CPI series for period-II, and the total and period-I series of China and Turkey, may be identified as bifractals instead of multifractals. Our results on $h(q)$ spectra apparently contradict the results of ref. \cite{Bolgorian11}, where it was possible to remove the multifractality of the original series by random shuffling. In a similar analysis by Ghosh et al. \cite{Ghosh12} a $q$-dependent $h(q)$ spectrum is observed for the shuffled series of gold indices. But one should keep in mind that the series time period in those analysis are different, and the recent rapid changing values of gold indices may moderately change the results. In our analysis the surrogate data generated by the AAFT procedure also result in $h(q)$ spectra that show significant amount of multifractality. The conjectures will be more transparent when we will discuss the singularity spectrum. It is to be noted that, in order to minimize the statistical error in the multifractal variables corresponding to the shuffled/surrogated series, we take an average of $h(q)$ values over 10 shuffled/surrogated series corresponding to each of the original series.

The multifractal $\tau(q)$ exponents are plotted against $q$ in Fig.~\ref{Fig4} for all the series under consideration. The lines in this figure also represent the binomial series, yielding Eqn.~\eqref{eq:tauq}, with the same sets of parameters that are used optimiz the $h(q)$ versus $q$ data and the parameters are given in Table \ref{Table1}. As already mentioned, for a monofractal series $\tau(q)$ is linear with $q$, and any kind of nontrivial dynamics present in the data is reflected in the form of a nonlinear variation of $\tau(q)$ with $q$. The signature of multifractality in all the original series is clearly visible from their nonlinear variation against of $\tau(q)$ order number $q$. In some cases the shuffled series as well as the surrogated series also show multifractality as strong as the original ones. Note that the observable $\tau(q)$ is directly calculated from $h(q)$, therefore we do not put any additional emphasis on it. However, we would like to mention that the GBM model reproduces the empirical $\tau(q)$ spectra very well.

\begin{figure}[t]
\centering
\includegraphics[width=6.75in,height=7in]{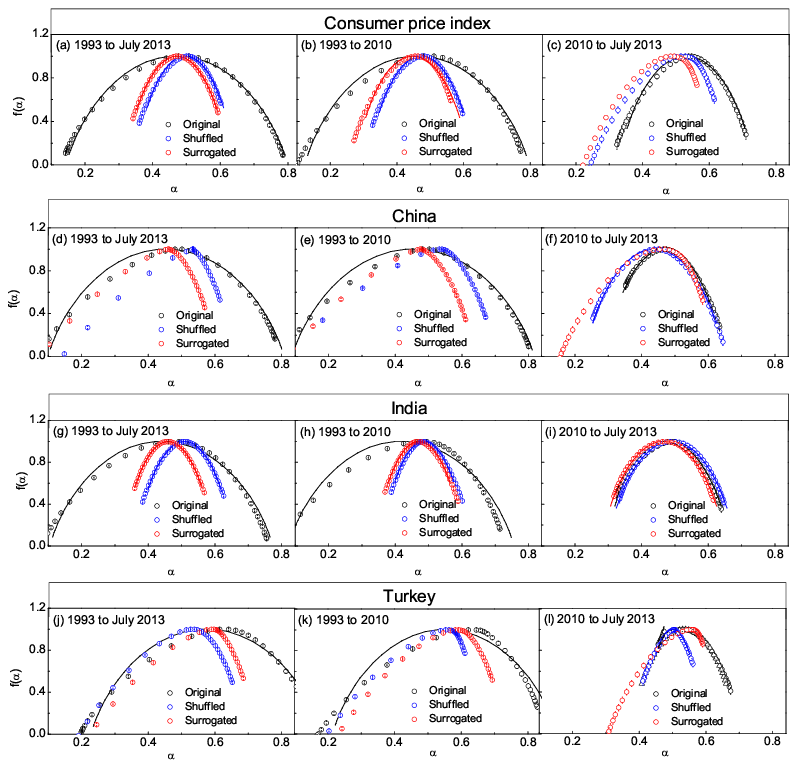}
\caption{The singularity spectra for the analyzed time series. The lines correspond to the GBM series prediction with the parameter values optimized for the $h(q)$ versus $q$ data points. The multifractal strength parameters values are given in Table \ref{Table1}.}
\label{Fig5}
\end{figure}
The most important results of a multifractal analysis is the singularity spectrum. To get a quantitative idea about the strength of multifractality, we calculate singularity spectra $f(\alpha)$ for the analyzed series and plot them in Fig.~\ref{Fig5} against the H\"{o}lder exponent $\alpha$. From the top to the bottom panel the diagrams represent, respectively the CPI time series, the price time series in China, India and Turkey. The solid lines in this figure also indicate the GBM series with the parameter values as given Table \ref{Table1}). As we know that the width of singularity spectrum is a measure of multifractality, the multifractal nature of all the original series is well reflected by the corresponding singularity spectrum. All the shuffled series within the time period 1993-July 2013 show much weaker multifractality than the corresponding original series. The same is true for the period-I series as well. But for the period-II shuffled series the singularity spectral widths are almost as equal as to their original counterparts. The observations here are consistent with our autocorrelation analysis that is, the time series for the period 2010--July 2013 does not capture long-range correlation, and hence the observed multifractal nature of these series arise purely from the probability distribution functions of their indices.    
The shuffled series generated spectra, as it is expected, are all centered around $\alpha \approx 0.5$. In some cases (especially Fig.~\ref{Fig5}(j) and (k)) the peak position of the $f(\alpha)$ spectrum is obtained at a slightly higher value of $\alpha$. A similar observation is also made in \cite{Grech10}, where it is shown that the peak position as well as the width of a multifractal spectrum may vary with its length along with its type. The surrogated series, irrespective of its time period, reproduce stable multifractal spectra but much narrower than the corresponding empirical ones. The strength parameter of multifractality ($\Delta \alpha$) is measured as the width of singularity spectrum at $f(\alpha)= 0$. Analytically that corresponds to $\Delta \alpha = (\ln a -\ln b)/\ln2 $. Using the ($a,~b$) parameter values we calculate the $\Delta \alpha$ values and are presented in the extreme right column of Table~\ref{Table1}. Here we will get a clear and quantitative difference among the various series studied. As can be seen in the table that the strength of multifractality for the original long and period-I series is always about two times that of the period-II series, while the difference between $\Delta \alpha$s of the original and the shuffled/surrogated series (whenever possible to fit the GBM model) for the series of period-II is marginal. The observations indicate that, except the market series in China and India for period-II, the observed multifractality in the analyzed gold price indices are partly due to the long-range time correlation and partly due to the fat-tailed distribution function of the values, and in the exceptional cases it is mainly due to the probability density function of the indices; for which we expect $\Delta \alpha$(original) $\approx \Delta\alpha$(shuffled). The surrogated data generated $f(\alpha)$ spectra in all cases but the series of Turkey are shifted towards the lower $\alpha$ side than their shuffled one, but they produce $\Delta\alpha$ value almost identical to that of the shuffled series. In case of the market of Turkey we find no change in the day-to-day market return index in several places, and some times the index remains constant over a wide section of the series. That might be a reason for the exceptional behavior of the market of Turkey.

\section{Conclusion}
Multifractal nature of (i) the consumer price index time series of gold and (ii) the gold market time series of China, Indian and Turkey during the period 1993--July 2013, has been investigated in terms of the multifractal detrended fluctuation analysis. For better understanding of the rapidly increasing and fluctuating market trend over the last few years we divide each series into two: first section from 1993 to 2010 (period-I) and second section from 2010 to July 2013 (period-II). That is, all together 12 original series, corresponding to each of them a randomly shuffled and a surrogated series, are analyzed. Multifractal observables, such as the generalized Hurst exponents ($h(q)$), multifractal exponent ($\tau(q)$) and singularity spectrum ($f(\alpha)$) for all the series, are extracted and are fitted (whenever possible) to the generalized binomial multifractal model series. The following conclusions can be drawn from the our analysis.

The MF-DFA fluctuation functions for all the analyzed time series nicely follow the scaling law \eqref{eq:Scaling}, as is expected for a multifractal series. The generalized Hurst exponent spectra corresponding to the original series are found to be order dependent and are fitted more or less by the GBM model. The shuffled series as well as the AAFT surrogated series also produce order dependent $h(q)$ spectra, but the order dependence is much weaker than the original series. The nature of the $h(q)$ spectra are consistent with the fact that the multifractality in the long-term CPI time series as well as in the market trend of India is generated from both the long-range temporal correlation and the probability distribution of the series values. In the case of the market series of China and Turkey the origin of multifractality is found to be the fat-tailed probability distribution function, and hence the binomial series \eqref{eq:binomial} cannot describe those series. The observed multifractality for period-II is mainly due to the probability density function. The $\tau(q)$ exponent diagrams reflect identical features of the data as the $h(q)$ diagrams do. The multifractal singularity spectra for all the original long series are significantly wider than the corresponding shuffled and/or surrogated spectra. For period-II the original and shuffled series produce almost identical singularity spectra that strengthen the observations of the $h(q)$ spectra.

It is obvious that the market series and the CPI series are highly correlated. Therefore, a cross-correlation study \cite{Horvatic11} between the individual market series and market-CPI series may reveal more insight into the gold market pattern. 
\bibliographystyle{plain}

\end{document}